# Microscopic Transport Analysis of Single Molecule Detection in MoS$_2$ Nanopore Membranes

Mingye Xiong[1], Michael Graf[2], Nagendra Athreya[1], Aleksandra Radenovic[2], and Jean-Pierre Leburton[1].

[1]Department of Electrical and Computer Engineering, and N. Holonyak Jr. Micro & Nanotechnology Laboratory, University of Illinois at Urbana-Champaign, Urbana, IL, USA, [2]Laboratory of Nanoscale Biology, Institute of Bioengineering, School of Engineering, Ecole Polytechnique Federal, Lausanne, Switzerland.

## Abstract

A microscopic physical analysis of the various resistive effects involved in the electronic detection of single biomolecules in a nanopore of a MoS$_2$ nanoribbon is presented. The analysis relies on a combined experimental-theoretical approach, where the variations of the transverse electronic current along the two-dimensional (2D) membrane due to the translocation of DNA and proteins molecules through the pore are compared with model calculations based on molecular dynamics (MD) and Boltzmann transport formalism for evaluating the membrane conductance. Our analysis that points to a self-consistent interaction among ions, charge carriers around the pore rim and biomolecules, emphasizes the effects of the electrolyte concentration, pore size, nanoribbon geometry, but also the doping polarity of the nanoribbon on the electrical sensitivity of the nanopore in detecting biomolecules, which agrees well with the experimental data.

The use of nanopores in solid-state membranes has been proven to be compelling in the quest for fast and accurate biomolecule of DNA and protein sensing without labeling or functionalization by monitoring electronic signals[1–6]. Among all the materials of choice for nanoporous membranes, two-dimensional (2D) solid-state materials such as graphene and transition metal dichalcogenides (TMD) stand out because of their high stability in different thermal and chemical conditions, but mostly for their ideal mono-atomic thickness comparable to the DNA base pair separation, which enables single nucleotide resolution.

Standard experiments on DNA translocation through narrow nanopores have already demonstrated DNA or protein sensing by detecting the usual blockade of the ionic current flowing between two electrolytic cells separated by a 2D membrane[7–9]. Although early studies of graphene nanopores showed partial success[10–13], they were impaired by the strong hydrophobic interaction between the membrane and nucleotides that results in severe sticking and clogging during DNA translocation[14,15]. However, they failed to expose the structural DNA details due to the low signal-to-noise ratio and the fast translocation speed. Alternatively, DNA translocation through the pore can be detected by the variations of the transverse electronic current flowing along a semiconducting TMD membranes such as molybdenum disulfide ($MoS_2$) that are less hydrophobic than graphene. While still under experimental realization[12], this approach has been predicted to result in larger and better resolved current signals with the potential for fast and reliable DNA sequencing[16].

Given the complexity of the nanopore systems involving the electric interaction between biomolecules, the ions in the electrolyte and the 2D solid-state membranes, three different interpretations have been proposed to explain the origin of the transverse electronic current variations in the 2D membranes caused by the presence of DNA in the nanopore. They are i) a capacitive effect between nanopore and solution [17], ii) a field-effect modulation of the carrier density in the membrane around the pore rim by the screened DNA charge[18], and iii) the bare interaction between nucleotides and the nanopore resulting in charge carriers density variations in the membrane[19–21]. Recently, a resistor circuit model was proposed to analyze the signal and noise of a 2D nanopore FET that excluded the latter interpretation iii) [22]. So far, no microscopic model has emerged to provide a coherent interpretation of the transverse current response to DNA translocation in solid-state nanoribbon nanopores. In particular, the recent experimental observation of transverse current variations dependent of the charge sign of the translocating biomolecules has not received a rigorous theoretical foundation[12].

In this work, we present a comprehensive physical analysis of the electronic current variation in a wide $MoS_2$ nanopore nanoribbon, which combines experimental transport characterization of the membrane with molecular dynamics and Poisson-Boltzmann modeling. The model takes into account the effect of biomolecules, electrolyte, and ion screening on the charge carriers in the nanoribbon to show that the DNA translocation induces significant potential varitions around the nanopore to affect the transverse current signal across the membrane. In this context, the role of the electrolyte concentration, pore size, and device geometry are critical in controlling the sensing sensitivity of the device, and underlines the self-consistent interaction among ions, biomolecule and charge carriers around the pore rim. Overall the model is consistent with the available experimental data.

The set-up of a MoS$_2$ nanoribbon nanopore is illustrated in figure 1(a), where a semiconducting MoS$_2$ nanoribbon of width $L_y$ and length $L_x$ is lying on a substrate (usually SiN$_x$), with a pore of diameter $d$ at the location $(x_0, y_0)$. Two electrodes are placed at the two ends of the MoS$_2$ nanoribbon, between which a voltage $V_{sd}$ is applied so that an electronic current flows along the nanoribbon (transverse current). In this set-up, the presence of the nanopore in the semiconducting nanoribbon acts as a scattering center for the electronic flow. Moreover, salt ions and DNA molecules passing through induce a time-varying electrostatic potential at the rim of the nanopore membrane, which modulates the strength of the scattering center, thereby causing transverse current variations. Thus, by virtue its sensitivity to the electrostatic properties of the nanopore, the scattered current ultimately detects the details of the DNA translocation.

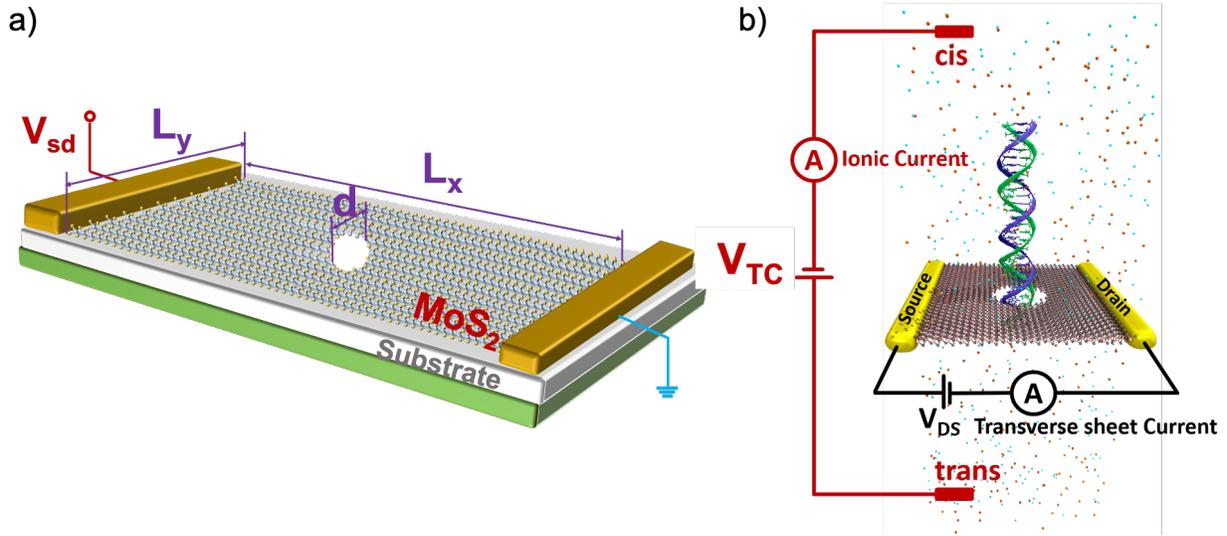

Figure 1. (a) Schematic (not to scale) of MoS$_2$ membrane device for transverse current measurement. Both ionic current and the transverse sheet current are measured. (b) Illustration of the system setup for dsDNA detection in electrolyte using both ionic current measurement and transverse current.

In order to investigate the influence of the scattering nanopore on the MoS$_2$ conducting electrons, we treat the electrostatic potential variations as a perturbation to the transverse current and use Fermi's Golden rule to assess its effect on the electronic conductivity. As the nanopore is 2-5 nm wide, which is much smaller than the width of the MoS$_2$ membrane ($L_y \sim$ 100-200 nm), one can model the perturbative nanopore as Dirac delta function potential

$$V(\vec{r}) = V_{tot}\delta(x - x_0)\delta(y - y_0)A_{np}, \qquad (1)$$

where $(x_0, y_0)$ is the position of the nanopore center, $A_{np}$ is the area of the nanopore, and $V_{tot}$ is the strength of the perturbation energy that consists in the combined electrostatic effects of DNA, ions and nanopore onto the conducting electrons. Because of the narrow width of the 2D MoS$_2$ membrane, we assume conducting electrons are 1D charge carriers confined along the $y$ direction, moving freely along the $x$ direction of the nanoribbon. Charge carriers are represented by their quantum mechanical wave function $\psi = Ce^{ikx}\sin\left(\frac{n\pi y}{L_y}\right)$ with the kinetic energy $E_{k,n,\alpha} = \frac{\hbar^2 k^2}{2m_\alpha} + E_n$, where $k$ is the electron wave vector in the $x$-direction, $E_n = \frac{\hbar^2 n^2 \pi^2}{2m_\alpha L_y^2}$ is the n$^{\text{th}}$ (n= 1,2…) sub-band energy in the $y$-direction and $m_\alpha$ is the effective mass of the electrons in the valley $\alpha$ in the MoS$_2$ band structure. Here $C = \sqrt{\frac{2}{L_x L_y}}$ is the normalization constant of the wavefunction. Thus, from Fermi's golden rule, the scattering rate of a single scattering from $(n, k)$ state to $(n', k')$ is

$$S(\alpha, k, k', n, n') = \frac{8\pi V_{tot}^2 A_{np}^2}{\hbar L_x^2 L_y^2} \sin^2\left(\frac{n\pi y_0}{L_y}\right) \sin^2\left(\frac{n'\pi y_0}{L_y}\right) \delta(E_{k',n',\alpha} - E_{k,n,\alpha}) \quad (2)$$

where $L_x$ is the length of the MoS$_2$ nanoribbon.

By summing over all initial $(n, k)$ and final $(n', k')$ states and using the linear response of the Boltzmann transport equation, we obtained the contribution to the membrane conductance solely caused by electron scattering by the nanopore delta potential,

$$G_p = \frac{F}{V_{tot}^2 A_{np}^2} \quad (3)$$

where $F$ is the form factor:

$$F_\alpha(L_y, y_0, E_f^\alpha) = -\frac{e^2}{2\pi} \frac{\hbar^2 L_y^2}{2\sqrt{2m_\alpha^3}} \sum_n \sin^{-2}\left(\frac{n\pi y_0}{L_y}\right) \int \frac{\sqrt{E-E_n}\, dE}{\sum_{n'} \frac{\sin^2\left(\frac{n'\pi y_0}{L_y}\right)}{k'(k,n,n')}} \frac{\partial f_\alpha(E)}{\partial E} \quad (4)$$

Here $f_\alpha(E)$ is the Fermi function of electrons in the MoS$_2$ valley $\alpha$ with the Fermi level $E_f^\alpha$ calculated with respect to the energy minimum of the valley. This form factor encapsulates the details of the nanoribbon geometry, pore position, material properties (such as carrier effective mass) and the carrier concentration. MoS$_2$ has been reported to consist of two valley minima in the conduction band corresponding to the K and Q valleys with slight energy difference $\Delta E_{QK} \approx 70$ meV and different effective mass $m_K = 0.5 m_0$ and $m_Q = 0.78 m_0$, so the overall form factor is $F(L_y, y_0) = F_K(L_y, y_0, E_f^K) + F_Q(L_y, y_0, E_f^Q)$. In figure 2, we display the form factor for different pore positions and various carrier concentrations, which shows significant variations with the pore position. As the conductance is inversely proportional to the form factor, it is seen that placing the nanopore at the center of the ribbon gives the highest conductance for the most sensitive bio-detection. However other positions at $y_0 = \frac{1}{5}L_y, \frac{2}{5}L_y, \frac{3}{5}L_y, \frac{4}{5}L_y$ are favorable as well. Furthermore, higher carrier concentration gives higher sensitivity. By taking into account intrinsic nanopore scattering mechanisms, the total nanoribbon conductance is given by Matthiessen's rule[23]

$$\frac{1}{G_{tot}} = \frac{1}{G_{ribbon}} + \frac{\gamma}{G_p} \quad (5)$$

where $\gamma = L_x/L_y$ is nanoribbon geometry aspect ratio, and $G_{ribbon}$ is the MoS$_2$ nanoribbon conductance in the absence of nanopore.

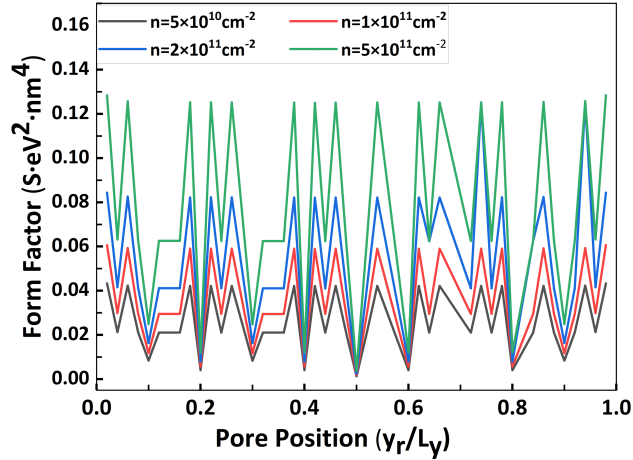

Figure 2. Form factor of the MoS$_2$ nanoribbon as calculated for various carrier concentrations and as a function of the pore position.

As mentioned above, the total perturbation potential $V_{tot}$ causing the variations of the electronic transverse current compared to those of a bare MoS$_2$ nanoribbon consists of all three effects

$$V_{tot} = V_{pore} + V_{electrolyte} + V_{DNA}. \qquad (6)$$

These effects are the perturbation due to the pore alone $V_{pore}$, the perturbation due to the presence of the electrolyte ions in the pore $V_{electrolyte}$ and that of the DNA in the pore $V_{DNA}$. In open pore, only the nanopore and electrolyte contribute to the total perturbation energy, $V_{tot}^{open} = V_{pore} + V_{electrolyte}^{open}$. Ideally, aside from the noise due to the ion stochastic transport, this perturbation remains unchanged throughout time. The transverse current due to these two factors sets the base current of a DNA translocation measurement. As a DNA molecule translocates through the nanopore, in addition to the DNA perturbation potential $V_{DNA}$ induced to the nanopore rim, both ions and counter-ions in the electrolyte re-arrange in the pore due to the DNA charge, thereby resulting in a change in the electrolyte perturbation $\Delta V_{electrolyte}$. Both the DNA presence and the redistribution of ions in the pore constitute the source of the transverse sensing signal.

In order to identify the contribution of each component of the total nanopore potential to the overall conductance variations, we perform experimental current measurements combined with molecular dynamics (MD) simulation on a nanoribbon nanopore connecting two electrolytic cells. Each cell consists of one pair of vertical electrodes driving the DNA translocation as well as the ionic current and one pair of lateral electrodes driving and measuring the transverse current along the nanoribbon (figure 1(b)). Here, the MoS$_2$ nanoribbon is $L_y = 500$ nm wide and $L_x = 2$ μm long. The nanopore with a diameter $d = 5.2$ nm is located at the center of the nanoribbon at $y_0/L_y = 0.5$.

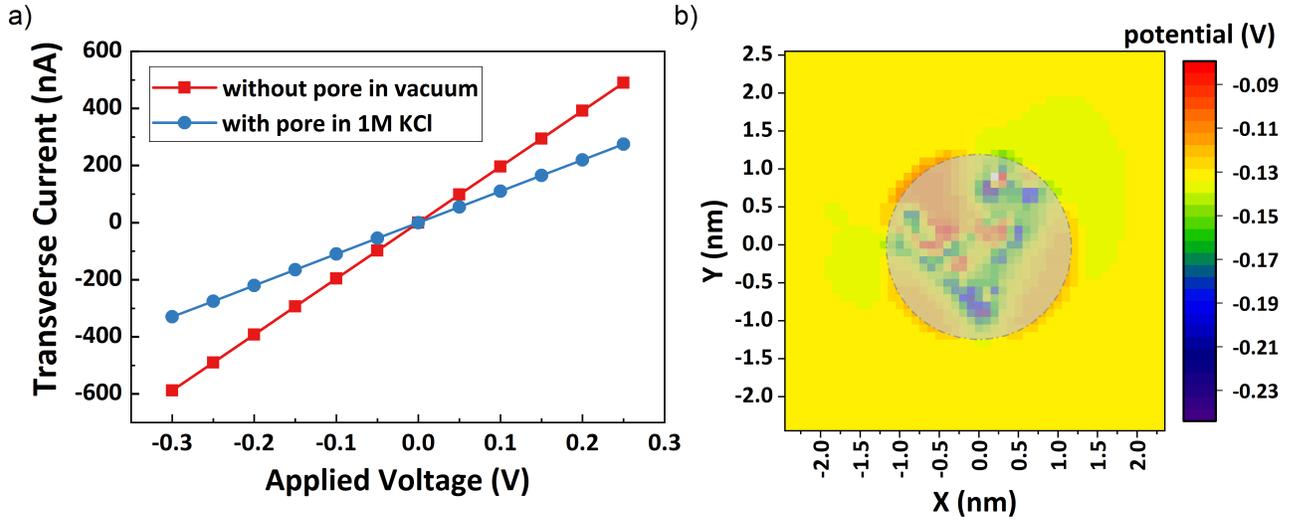

Figure 3 (a) I-V curves of the MoS$_2$ nanoribbon in vacuum before drilling the pore (red curve) and after pore drilling ( blue curve) in the center of the nanoribbon in a 1M KCl solution; (b) potential color plot in the pore and around the pore rim for a dsDNA translocation through a 2.6 nm pore in a MoS$_2$ membrane at $t = 200$ au in a 10mM KCl solution.  The grey shaded area represents the pore. The potential outside the pore is compiled for the calculation of electron scattering.

First, we focus on the transverse electronic conductance resulting from the presence of an open pore to extract the potential $V_{pore} + V_{electrolyte}$. For this purpose, we measure the conductance of a MoS$_2$ nanoribbon in a vacuum and the conductance of the same nanoribbon after drilling the nanopore in the presence of one mole KCl solution (figure 3(a)) from which we obtain the bare MoS$_2$ nanoribbon $G_{ribbon} = 1.96$ µS, and $G_{tot}^{op} = 1.1$ µS for the open pore measurement, yielding $G_p = 10$ µS for $\gamma = \frac{L_x}{L_y} = 4$. Recent experimental and theoretical works[24–29] have shown that electrons in a single layer MoS$_2$ are characterized by a mobility ranging from 10 to 50 cm$^2$V$^{-1}$s$^{-1}$, from which, based on the nanoribbon conductance, the carrier concentration varies between $0.61\times10^{11}$ cm$^{-2}$ and $3.1\times10^{11}$ cm$^{-2}$ corresponding to a Fermi level between 67 meV and 46 meV below the conduction band edge. Based on these values one can calculate the form factor ranging between $F = 1.40\times10^{-3}$ S·eV²nm⁴ and $F = 3.04\times10^{-3}$ S·eV²nm⁴. By combining equations (3) and (5) into $V_{tot} = \sqrt{\left(\frac{1}{G_{tot}} - \frac{1}{G_{ribbon}}\right)\frac{F}{\gamma}\frac{1}{A_{np}^2}}$, one can estimate the open pore potential

$$V_{tot}^{op} = V_{pore} + V_{electrolyte}^{op} \qquad (7)$$

to vary between 0.56 eV and 0.82 eV. In addition to the nanopore blocking the electrons, the pore drilling process will also cause certain degradation of the MoS$_2$, which also contributes to decrease in conductance. This degradation effect is also included in the $V_{pore}$ potential.

Second, in order to assess the magnitude of the electric potential energy $V_{DNA}$ generated around the nanopore during a DNA translocation event, we make use of Molecular Dynamics (MD) to simulate the electrophoresis of DNA molecules and extract the DNA charge distributions during the translocation. We then use a self-consistent Poisson-Boltzmann scheme to calculate the re-distribution of ion charges caused by the DNA presence in the nanopore. The details of MD simulation and self-consistent Poisson-Boltzmann scheme are described in the Method section. In figure 3(b), we display the colormap of the electric potential that exhibits significant negative values due to the negative charge of the DNA backbones inside a nanopore of 2.6 nm diameter in 1 M KCl electrolyte. From the color code, one can clearly see that the negative potential extends inside the MoS$_2$ membrane, which adds to the electron scattering by the pore and the ions, and in turn modifies the transverse electronic current. After compiling the values of the DNA potential variations around the pore for each time frame of the DNA trajectory obtained by MD, one obtains the potential variation of the DNA translocation event as a function of time. Figure 4(a) (right scale) compares the traces of the potential variations around a 2.6 nm wide nanopore for two KCl molarities, where it is seen the potential variations arising from DNA translocation in a 10 mM KCl solution result in a ~20 mV amplitude change on the pore rim, while in a 1 M KCl solution, the potential variation amplitude is smaller, and reaches barely ~6 mV at its maximum value. This amplitude difference is due to the ion screening effect characterized by the Debye length, which in water at room temperature is $\lambda_d = \frac{0.304}{\sqrt{I(M)}}$ (in nm), where $I$ is the ionic molar concentration (M). Hence, in 10 mM KCl solution, $\lambda_d \sim 3$ nm, whereas in 1 M KCl solution, it is one tenth that length, i.e. ~0.3 nm. This larger electrical sensitivity of MoS$_2$ nanopores in lower electrolytic concentration is in qualitative agreement with recent experimental data. The simulations also show that the high-frequency fluctuations of the potential variations (in the order of a few mV for the 10 mM KCl electrolyte) are caused by mainly the stochastic DNA conformational variations during the translocation event, as well as the nucleotide difference along the DNA strand and the redistribution of ions due to the DNA charge in the pore $\Delta V_{electrolyte}$. Thus, MoS$_2$ in a lower concentration electrolyte can sense more of the potential change coming from DNA translocation.

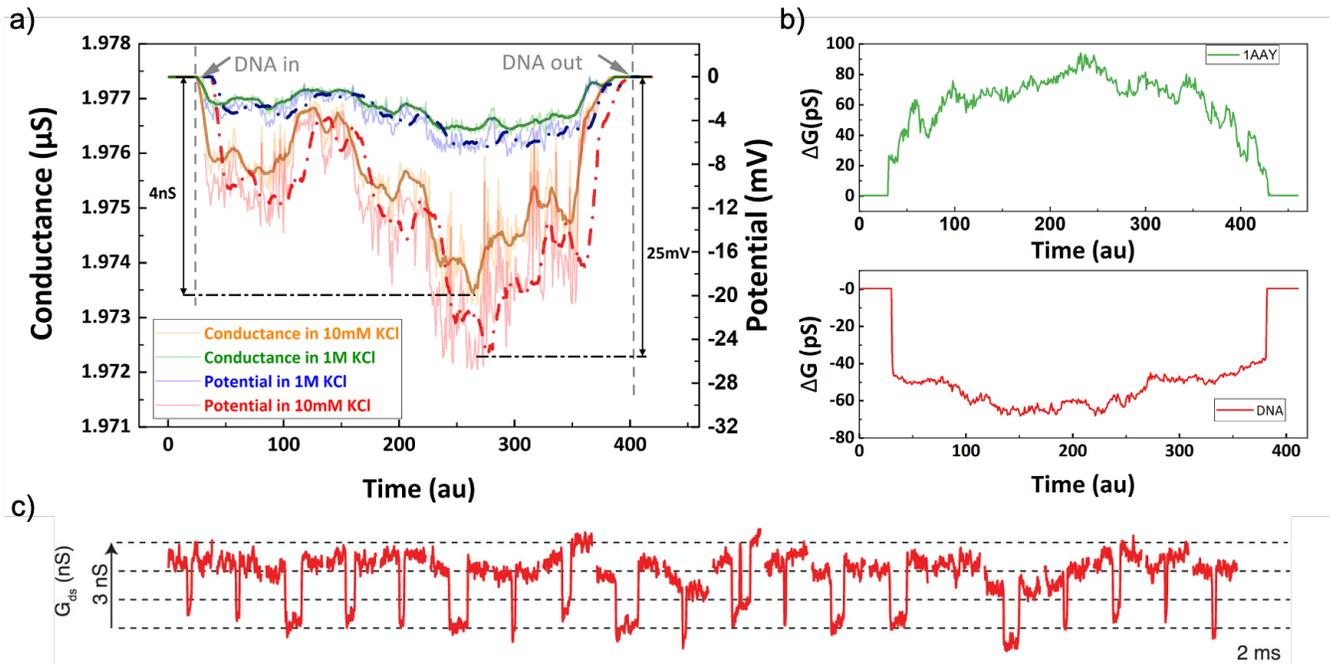

Figure 4. (a) Potential and transverse conductance variation during a 30bp dsDNA translocation through a 2.6 nm MoS$_2$ nanopore in a 10 mM KCl and 1 M KCl solutions: the conductance in 10 mM KCl shows a clear signal of translocation which agrees well with the experimental results displayed in (c) below ; (b) Transverse conductance variation during the translocation of a positively charged protein 1AAY (top) and negatively charged dsDNA (bottom) through 5.2nm MoS$_2$ nanoribbon nanopore; (c) Experimental traces of multiple 1kb dsDNA translocating through a 2.5 nm pore in a 10 mM-1 M (cis-trans) KCl solution. The magnitude of the dips is in agreement with the theoretical results displayed in figure (a). Figure (c) reprinted from *Transverse Detection of DNA Using a MoS2 Nanopore,* by Graf, M., et al. Copyright 2019 *Nano Letters*.

We now proceed to calculate the transverse current by using our transport model. As shown in figure 4(a) (left scale), the transverse conductance for a dsDNA translocation event in 10 mM KCL varies with an amplitude of ~4 nS. This agrees well with the experimental results as shown in figure 4(c)[12]. As a consequence of the reduced potential variations in high electrolytic concentration, the conductance variations in 1 M KCl are below 1 nS, which is too small to be observed experimentally. Because these electric variation effects are relatively small compared to the presence of the pore itself and the ions in the open pore, i.e.

$$V_{DNA} + \Delta V_{electrolyte} = \delta V \ll V_{tot}^{op} = V_{pore} + V_{electrolyte}^{op}$$

This results in a relatively small conductance change $\delta G$ in the MoS$_2$ ribbon, so that

$$\frac{1}{G_{tot}^{op} + \delta G} = \frac{1}{G_{ribbon}} + \frac{\gamma (V_{tot}^{op} + \delta V)^2 A_{np}^2}{F} \tag{8}$$

where

$$\frac{1}{G_{tot}^{op}} = \frac{1}{G_{ribbon}} + \frac{\gamma (V_{tot}^{op})^2 A_{np}^2}{F} \tag{9}$$

From these equations (8) and (9) one gets the linear relation between $\delta G$ and $\delta V$, i.e.

$$\delta G = -\frac{2\gamma A_{np}^2 V_{tot}^{op} G_{tot}^{op\,2}}{F} \delta V \tag{10}$$

In equation (10), one notices that when $V_{tot}^{op}$ and $\delta V$ are of the same sign, the conductance correction due to DNA translocation in the nanopore is negative as shown in figure 4(b). Here the negative charges on the DNA backbones induce a negative electric potential on the nanopore rim, which increases the electron perturbation energy in the MoS$_2$ nanoribbon, thereby decreasing the electronic conductance. This sign dependence of the perturbation energy on the charge of the translocating biomolecule is consistent with recent experiments that show an increase of the transverse conductance during translocation of positively charged polylysine protein[12]. For this reason, we further performed the same kind of simulation of a translocating positively charged 1AAY protein, for which owing to the protein size, we enlarge the pore to a 5.2 nm diameter. Figure 4(b) displays the transverse conductance variations for both translocating bio-molecules, i.e. 1AAY protein and DNA, showing opposite variations with the former conductance increasing and the latter decreasing, as expected. In this comparison, it should be emphasized that in the case of p-type $MoS_2$, i.e. the charge carriers are holes, $V_{tot}^{op}$ will be inverted and the conductance variations will also be inverted for both types of bio-molecules, so that the sign of the conductance variation is not only due to the charge of the bio-molecule, DNA or proteins, which determines the sign of $\delta V$, but also the membrane type, as well as the nanopore charge that determines $V_{tot}^{op}$.

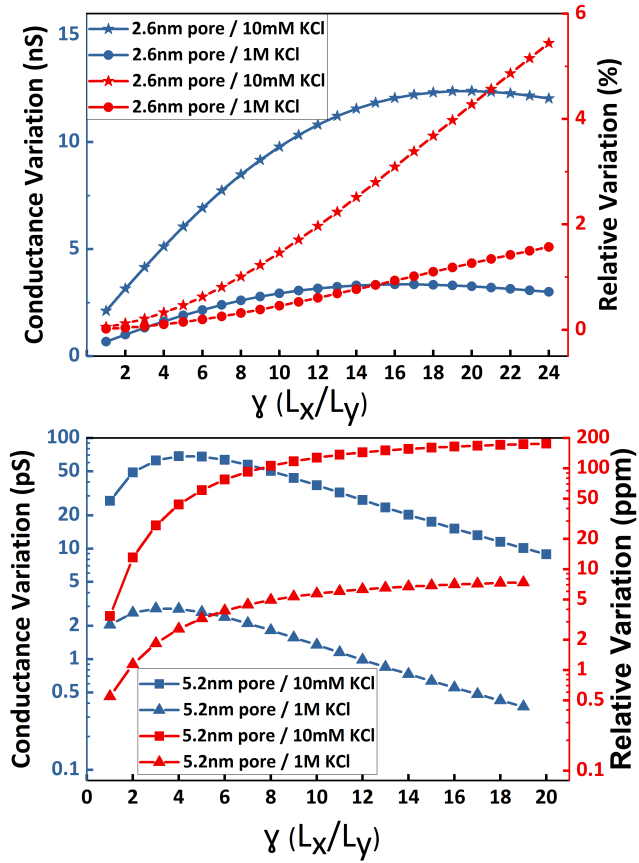

Figure 5. Absolute and relative conductance variation of a dsDNA translocation event through a 2.6 nm pore (top) and 5.2 nm pore (bottom) as a function of the MoS$_2$ nanoribbon geometric aspect ratio ($\gamma = L_x/L_y$).

As the electronic conductance variation $\delta G$ captures the features of the DNA translocation signal, by amplifying its value one enhances the sensitivity of the MoS$_2$ nanopore biosensor. Among the different factors shaping $\delta G$, the nanoribbon geometry depicted by the geometric aspect ratio $\gamma = L_x/L_y$ is critical. As shown in equation (10), $\delta G$ is proportional to $\gamma$ and inversely proportional to $F$, and thereby inversely proportional to $L_y$. In figure 5 (top), we display the absolute and relative conductance variation $\delta G$ (left scale) and $\delta G/G$ (right scale), respectively, as a function of $\gamma$, where one can see that $\delta G$ exhibits a maximum at $\gamma = 16$ for 1 M KCl and at $\gamma = 20$ for 10 mM KCl. This behavior is the consequence of the decreasing dependence of the open pore conductance $G_{tot}^{op}$ on $\gamma$, as clearly seen in the monotonic variation of $\delta G/G$ with the geometric aspect ratio. The phenomenon dominates with 5.2 nm pore, as shown in figure 5 (bottom) that displays the maximum of $\delta G \approx 2.9$ pS at $\gamma = 3$ in 1 M KCl and $\delta G \approx 68$ pS at $\gamma = 4$ in 10 mM KCl, and decreases drastically as $\gamma$ further increases (the width $L_y$ further shrinks).

Another key factor that affects the sensitivity is the pore size, which can be seen by comparing the conductance signal in a 2 nm (figure 4(a)) and a 5 nm pore (figure 4(b)), where one observes a sensitivity difference of about 60 times larger in the former. This large difference results from the fact that a wider pore experiences reduced DNA perturbation variations caused by the screening of the ions determined by the Debye length, $\lambda_d$. However, in equation (10), one also sees that as the pore size ($A_{np}$) increases, the conductance variation induced by the biomolecule translocation increases. For the two simulated nanopore sizes, one observes that the decrease in $\delta V$ plays a more significant role, and thus smaller nanopore gives a higher sensitivity. In general, however, sensitivity optimization is a delicate balance between the two factors $A_{np}$ and $\delta V$.

**Conclusion**

In this paper, we use a combined experimental-theoretical approach to analyze in detail the biosensing process in a nanopore by transverse electronic current variations in a MoS$_2$ membrane nanoribbon. Our microscopic analysis offers a direct connection between the transverse current response to biomolecule translocations in the nanopore and the different components of the electrical resistance of the membrane, i.e. electrolyte, open pore, and DNA motion. In particular, our model based on the Boltzmann transport formalism emphasizes not only the effects of the electrolyte concentration, pore size, and the nanoribbon geometric aspect ratio $\gamma = L_x/L_y$, but also that of the doping polarity of the nanoribbon, on the electrical sensitivity of the nanopore. These effects underline the microscopic interaction among the three charged systems at play in the detection mechanism, i.e. ions, biomolecules and charge carriers in the membrane. As a result, we note that the transverse electronic conductance response to dsDNA translocation in a 2.6 nm pore agrees well with experimental results. Finally, let us point out that our model for nanopore biosensing can easily be extended to any 2D material other than MoS$_2$ by accounting for the details of the material electronic band structure.

**Methods**

**Molecular Dynamics (MD) Simulations**

In the first step of the simulations, the intended nanopore system was built by using all-atom molecular dynamics (MD) software complemented by the Visual Molecular Dynamics (VMD) software for system analysis [30], while 9 nm x 9 nm $MoS_2$ membranes were constructed by plugging material parameters in the VMD. In particular, Lennard-Jones parameters from Stewart et al.[31] were incorporated for the atoms in $MoS_2$ monolayer and all the atoms in the $MoS_2$ membrane were fixed to their initial positions to avoid the drifting of the membrane during the simulations. Then, nanopores in the membrane were created by manually removing relevant atoms. The dsDNA structure was obtained from the 3D-DART web server[32] and was described using CHARMM27 force fields[33]. For illustrating the device performance on positively charged molecule, Zif268 protein-DNA complex (PDB code: 1AAY), a DNA binding protein which is inherently positively charged, was considered[34]. The protein was described using CHARMM22 force field with CMAP corrections[35]. To shorten the simulation time, the DNA molecule or protein was placed above the nanopore to begin the translocation. The whole system is solvated in a water box and ions (potassium and chlorine) are randomly placed in box to reach a concentration of 10 mM or 1 M. An electric field is applied perpendicular to the nanopore membrane to drive the biomolecule through the pore. The MD simulations are performed using NAMD 2.13[36]. The systems are maintained at 300 K using a Langevin thermostat. Periodic boundary conditions are employed in all directions. Time steps of 2 fs along with a particle Mesh Ewald[37] were used to treat long-range electrostatics. All systems are minimized for 5000 steps, followed by a 600 ps equilibration as an NPT ensemble. The systems are further equilibrated as an NVT ensemble for another 2 ns before the electric field is applied. The trajectories of all atoms in the system are recorded at every 5000 steps until the DNA or DNA-protein complex is completely translocated.

**Potential Calculations**

For each frame in the MD trajectory, electrostatic potential, $\varphi(r)$, is calculated using Poisson Boltzmann's equation shown in equation 1.

$$\nabla \cdot [\epsilon(r)\nabla\varphi(r)] = -e\,[C_{K^+}(r) - C_{Cl^-}(r)] - \rho_{DNA/protein}(r) - \rho_{semiconductor}(r) \quad 1$$

where $\rho_{DNA/protein}(r)$ is the charge due to DNA or protein, $\rho_{semiconductor}(r)$ is the mobile charges in the MoS$_2$ layer, $\epsilon(r)$ being the local permittivity, and $C_{K^+}(r)$ and $C_{Cl^-}(r)$ are the local electrolyte concentrations of $K^+$ and $Cl^-$ that obey Poisson-Boltzmann statistics given by the following equations.

$$C_{K^+}(r) = C_0 exp\left[\frac{-e\varphi(r)}{k_B T}\right] \quad 2$$

$$C_{Cl^-}(r) = C_0 exp\left[\frac{e\varphi(r)}{k_B T}\right] \quad 3$$

Here, $C_0$ is the nominal concentration in the solution, which is set to 10 mM or 1 M. The above two equations are solved numerically until the convergence criterion is met. A detailed description of charge distributions is given by Girdhar et al.[38].

**Acknowledgements**

M.X., N.A, and J-P.L. are indebted to Oxford Nanopore Technology and the Illinois-Proof-of-Concept (i-POC) program for supporting this study, and gratefully acknowledge supercomputer time provided through the Extreme Science and Engineering Discovery Environment (XSEDE) Grant TG-MCB170052. M.G. and A.R. acknowledge support from the Swiss National Science Foundation (SNSF) Consolidator grant (BIONIC BSCGI0_157802) and CCMX project ("Large Area Growth of 2D Materials for Device Integration").